\documentclass[aps,twocolumn]{revtex4}
\usepackage[normalem]{ulem}
\usepackage{bm}
\usepackage[usenames, dvipsnames]{color}
\definecolor{gre}{rgb}{0.0, 0.42, 0.24}
\usepackage{graphicx,amsmath}
\usepackage[colorlinks]{hyperref}
\hypersetup{colorlinks,citecolor=red,linkcolor=blue,urlcolor=blue}

\begin{document}

\title{On the discreet spectrum of fractional quantum hydrogen atom in two dimensions.}

\author{V. A. Stephanovich}\affiliation{Institute of Physics, Opole
University,\\Oleska 48, 45-052, Opole, Poland}

\date{\today}

\begin{abstract}
We consider a fractional generalization of two-dimensional (2D) quantum-mechanical Kepler problem corresponding to 2D hydrogen atom. Our main finding is that the solution for discreet spectrum exists only for $\mu>1$ (more specifically $1 < \mu \leq 2$, where $\mu=2$ corresponds to "ordinary" 2D hydrogenic problem), where $\mu$ is the L\'evy index. We show also that in fractional 2D hydrogen atom, the orbital momentum degeneracy is lifted so that its energy starts to depend not only on principal quantum number $n$ but also on orbital $m$. To solve the spectral problem, we pass to the momentum representation, where we apply the variational method. This permits to obtain approximate analytical expressions for eigenvalues and eigenfunctions with very good accuracy. Latter fact has been checked by numerical solution of the problem. We also found the new integral representation (in terms of complete elliptic integrals) of Schr\"odinger equation for fractional hydrogen atom in momentum space. We point to the realistic physical systems like bulk semiconductors as well as their heterostructures, where obtained results can be used.
\end{abstract}
\maketitle

\section{Introduction}

The excitons in two-dimensional (2D) semiconductor structures, especially in the presence of disorder,  is a topic of intensive experimental and theoretical studies \cite{bvb,as1,as2,ohm,du1}. Such studies become particularly useful for semiconductor quantum wells and interfaces that embrace number of functionalities like photovoltaic cells and nanolasers \cite{cells,nanolas}. It is well known (see, e.g. \cite{anselm,ash}), that the exciton is a physical realization of quantum Kepler problem or hydrogen atom. It have been shown recently \cite{she1,she2} that adding the spin-orbit interaction in Rashba form \cite{br} generates the chaotic motion in the above excitons, which can degrade the performance of the exciton-based devices. On the other hand, the disorder can also adversely influence such devices functionality. The convenient tool to account for disorder in such a quantum system is to substitute the ordinary Laplacian in the Schr\"odinger equation by the fractional one. In other words, the problem of fractional 2D hydrogen atom can serve as a model example of an exciton in disordered semiconductor quantum well and/or interface. 

The definition of the fractional Laplacian in two spatial dimensions reads:
\begin{eqnarray}
-|\Delta|^{\mu/2}f({\bf x})=A_\mu \int \frac{f({\bf u})-f({\bf x})}{|{\bf u}-{\bf x}|^{\mu +2}}d^2u, \label{omm2} \\
A_\mu=\frac{2^\mu\Gamma\left(\frac{\mu+2}{2}\right)}{\pi|\Gamma(-\mu/2)|}. \label{omm3}
\end{eqnarray} 
Here $0<\mu<2$ is L\'evy index and $\Gamma(x)$ is $\Gamma$ - function \cite{abr}. With this definition at hand, the fractional Schr\"odinger equation for 2D hydrogen atom reads
\begin{equation}\label{pmm1}
-|\Delta|^{\mu/2}\Psi_{nm\mu}({\bf r})-\frac{2}{r}\Psi_{nm\mu}({\bf r})=E_{nm\mu}\Psi({\bf r}),
\end{equation}
where ${\bf r}$ is a two-dimensional vector, indices $n$ and $m$ denote, respectively, the principal and orbital quantum numbers, which are different for any specific $\mu$ value. Below we shall see that for $\mu<2$ the 
orbital degeneracy is lifted and that is the reason why the eigenenergy $E$ has now two subscripts. 
Here we use modified (for the fractional case $\mu<2$) Rydberg units \cite{laskin}, i.e. we measure the energy $E$ and coordinates ${\mathbf r}$ in the units
\begin{equation}\label{pmm2}
E_{0\mu}=\left(\frac{\beta}{2\hbar}\right)^{\frac{\mu}{\mu-1}}D_\mu^{-\frac{1}{\mu-1}},\ r_{0\mu}=\left(\frac{2\hbar ^\mu D_\mu}{\beta}\right)^{\frac{1}{\mu-1}}
\end{equation}
respectively. Here $\beta$ is a coefficient in front of (dimensional) Coulomb potential:
\begin{equation}\label{kul}
U(r)=-\frac{\beta}{r},
\end{equation}
$D_\mu$ is a mass term \cite{laskin}. At $\mu=2$ $D_2\equiv \frac{1}{2m}$ ($m$ is a real physical mass) and we have from \eqref{pmm2} $E_{0,\mu=2}=\frac{m\beta^2}{2\hbar^2}$, $r_{0,\mu=2}=\frac{\hbar^2}{m\beta}$, i.e. standard Rydberg units. We pay attention to one more fact. Namely, at $\mu=1$ both quantities $E_{0,\mu=1}$ and $r_{0,\mu=1}$ are divergent. Below we will see that this reflects the actual situation with fractional 2D hydrogenic problem, i.e. that discreet spectrum of the problem exists for $\mu>1$ only and at $\mu \to 1$ the whole spectrum goes to minus infinity.

It is well-known that integral \eqref{omm2} exists only in the sense of Cauchy principal value. This already complicates the solutions of spectral problems for pseudo-differential equations involving this operator and Eq. \eqref{pmm1} in particular. Below we shall see that it is much more profitable to pass to the momentum space as the operator \eqref{omm2} in this space renders simply as $-k^\mu$ ($k\equiv |{\bf k}|$). Although the potential term \eqref{kul} converts to the integral in momentum space, this integral proves to be much easier to handle then initial one \eqref{omm2}. Moreover, below we show that we can do angular integration without expansion over spherical harmonics (which is customary in quantum hydrogenic problems, see \cite{pp} and references therein) and by that virtue derive very useful representation of the equation \eqref{pmm1} in momentum space. Latter representation will permit to solve the corresponding fractional Schr\"odinger equation both numerically and variationally. To find the eigenfunctions in the ${\mathbf r}$ space, we perform inverse Fourier transformation. 

In the present paper we focus on the discreet spectrum of 2D fractional quantum Kepler problem. We will show that the solution to this problem exists only for L\'evy index $1<\mu \leq 2$, which is already surprising fact. We will also show that for $\mu \neq 2$ the well-known orbital degeneracy for quantum hydrogenic problem (see, e.g. \cite{land3} for 3D case and \cite{pp,kit} for 2D one) is lifted. That is to say, that imperfections, leading to fractional derivatives introduction to the problem, lower its symmetry, thus lifting the orbital degeneracy. 

\section{The spectral problem in momentum space}
%\section{General formalism}
We convert the problem \eqref{pmm1} to the momentum space by means of symmetric Fourier transform
\begin{subequations}
\begin{eqnarray}
f({\bf k})=\frac{1}{2\pi}\int e^{i{\bf k}{\bf r}}f({\bf r})d^2r,\label{ftran1} \\
f({\bf r})=\frac{1}{2\pi}\int e^{-i{\bf k}{\bf r}}f({\bf k})d^2k. \label{ftran2}
\end{eqnarray}
\end{subequations}
Such symmetric transform is convenient for our purposes as it conserves the normalization of wave function $\Psi$
both in coordinate and momentum spaces
\begin{equation}\label{qmm1}
\int |\Psi(\mathbf{r})|^2d^2r=1,\ \int |\Psi(\mathbf{k})|^2d^2k=1.
\end{equation}  
We apply Fourier transformation \eqref{ftran1} to both parts of equation \eqref{pmm1} to obtain $(E-k^\mu)\Psi({\bf k})-W({\bf k})=0$ (we suppress subscripts for a moment), where
\begin{eqnarray}
W({\bf k})=\frac{1}{2\pi}\int V(r)\Psi({\bf r})e^{i{\bf kr}}d^2r  \nonumber \\
\equiv \frac{1}{(2\pi)^2}\int V(|{\bf k}-{\bf k'}|)  \psi({\bf k'}) d^2k',\label{qmm2}
\end{eqnarray}
where 
\begin{equation} \label{qmm3}
V(r)=-\frac 2r,\ V(q)=-\frac{4\pi}{q},
\end{equation}
($q\equiv |{\bf q}|$) are, respectively, Coulomb potential and its Fourier image. Plugging everything together, we arrive at following final form of the equation \eqref{pmm1} in momentum representation
\begin{equation}\label{momm}
\left(k^\mu+k_0^\mu \right)\Psi({\bf k})-\frac{1}{\pi}\int d^2k' \frac{\Psi({\bf k'})}{|{\bf k}-{\bf k'}|}=0,
\end{equation}
where we once more suppress the indices $nm\mu$. Here, similar to the case of ordinary 2D hydrogen atom (see, e.g. \cite{pp}), we denote
\begin{equation} \label{ek0}
E=-k_0^\mu.
\end{equation}
It is easy to see, that at $\mu=2$ the equation \eqref{momm} yields well-known form for ${\bf k}$ - space representation of "ordinary" (i.e. with normal Laplacian, corresponding to $\mu=2$) 2D hydrogen atom \cite{pp}. 

As Coulomb potential \eqref{kul} is central, the solutions are invariant under the rotation around $z$ axis. In other words, $z$ - projection of the angular momentum is conserved (which is sufficient in 2D) and we can separate the radial (in ${\bf k}$ space) and angular variables. This implies that the solution can be represented in the form
\begin{equation}\label{angvar}
\Psi_{nm\mu}({\bf k})=\psi_{nm\mu}(k)e^{im\varphi},
\end{equation}
where $k=|{\bf k}|$ is vector ${\bf k}$ modulus and $\varphi$ is its asimuth angle. As usual in hydrogenic problems, radial functions $\psi_{nm\mu}(k)$ are real.  With respect to \eqref{qmm1}, the normalization condition for radial functions becomes

 \begin{equation}\label{ormm}
 2\pi \int_0^\infty \psi_{nm\mu}^2(k)kdk=1.
 \end{equation}

Substitution of \eqref{angvar} into the Schr\"odinger equation \eqref{momm} yields
\begin{eqnarray}
\left(k^\mu+k_0^\mu \right)\psi_{nm\mu}(k)e^{im\varphi}-\frac{1}{\pi}\int_0^{2\pi} d\varphi' \nonumber \\
\times \int_0^\infty  \frac{\psi_{nm\mu}(k')e^{im\varphi'}k'dk'}{\sqrt{k^2+k'^2-2kk'\cos(\varphi-\varphi')}}=0. \label{dox8}
\end{eqnarray}
Performing the substitution $\varphi - \varphi' =t$ in the angular integral \eqref{dox8}, we transform this equation to the form
\begin{eqnarray}
\left(k^\mu+k_0^\mu \right)\psi_{nm\mu}(k)e^{im\varphi}\nonumber \\
+\frac{e^{im\varphi}}{\pi}\int_0^\infty I_m(k,k')\psi_{nm\mu}(k')k'dk'=0, \label{dox9}
\end{eqnarray}
which, after cancellation of the factor $e^{im\varphi}$, gives the following equation for radial component
\begin{equation}
\left(k^\mu+k_0^\mu \right)\psi_{nm\mu}(k)
+\frac{1}{\pi}\int_0^\infty I_m(k,k')\psi_{nm\mu}(k')k'dk'=0, \label{dox10}
\end{equation}
where for clarity we suppress the subscripts $mn\mu$ in the parameter $k_0$ and
\begin{equation}\label{dox11}
I_m(k,k')=\int_\varphi^{\varphi-2\pi} \frac{e^{-imt}dt}{\sqrt{k^2+k'^2-2kk'\cos t}}.
\end{equation}
The possibility to cancel the exponent $e^{im\varphi}$ in \eqref{dox9} is one more demonstration of the correctness of {\em{ansats}} \eqref{angvar}. Note that at $\mu=2$, the Scr\"odinger equation \eqref{momm} is usually solved by Fock's method \cite{fock} (see also Ref. \cite{levy1} for discussion of $d=q+2$ dimensional case) of stereographic projection \cite{pp}. In this method, the angular integrals are not calculated explicitly; rather, the spherical harmonics expansion of the wave function is used, giving the exact solution of the problem for $\mu=2$. Our analysis shows that for $\mu<2$ such stereografic projection is impossible for general $\mu$, which is continuous variable. That's why here we adopt a different strategy. Namely, we calculate the integrals $I_m(k,k')$ analytically, which permits to solve the problem variationally, obtaining the approximate values of eigenenergies for each $\mu$ as well as the expressions for corresponding wave functions. The variational results will then be checked numerically. 

Although it is impossible to calculate analytically $I_m$ for general $m$, it is possible to do that for each specific $m$. This calculation shows  explicitly that $I_m(k,k')$ does not depend on $\varphi$, which once more is the consequence of angular momentum conservation and hence of the possibility to separate the variables \eqref{angvar}. 

We have for $m=0$
 \begin{eqnarray}
  I_0(k,k')=\int_\varphi^{\varphi-2\pi} \frac{dt}{\sqrt{k^2+k'^2-2kk'\cos t}}=\nonumber \\
  = -\frac{4}{k+k'}K\left(\frac{4kk'}{(k+k')^2} \right), \label{dox12}
 \end{eqnarray}
where $K(m)$ is a complete elliptic integral of the first kind \cite{abr}. The details of derivation of \eqref{dox12} are listed in the  Appendix \ref{intt}. Note that at $k=k'$ the integral $K(...)$ in \eqref{dox12} is divergent as its argument equals to 1. This reflects the spherical harmonic series divergence at $k=k'$ both from the side $k<k'$ and $k>k'$. Because of weak (logarithmic) character of divergence, it is well compensated by other parts of corresponding integrands. This once more shows the usefulness of the closed form analytical representation for $I_m$ as compared to spherical harmonics expansion.
 
The expressions for the integrals $I_1(k,k')$ and $I_2(k,k')$ can be obtained in a similar way, see Appendix \ref{intt} for details. After lengthy calculations we arrive at
\begin{eqnarray}
&&I_1(k,k')=\int_\varphi^{\varphi-2\pi} \frac{e^{-it}dt}{\sqrt{k^2+k'^2-2kk'\cos t}}=\nonumber \\
&&-\frac{2}{kk'(k+k')}\bigg[(k^2+k'^2)K-(k+k')^2E\bigg]. \label{ii1} \\
&&I_2(k,k')=\int_\varphi^{\varphi-2\pi} \frac{e^{-2it}dt}{\sqrt{k^2+k'^2-2kk'\cos t}}=\nonumber \\
&&-\frac{4}{3k^2k'^2(k+k')} \bigg[(k^4+k'^4+k^2k'^2)K-\nonumber \\
&&-(k^2+k'^2)(k+k')^2E\bigg], \label{ii2}
\end{eqnarray}
where
\[
K\equiv K\left(\frac{4kk'}{(k+k')^2} \right),\ E\equiv E\left(\frac{4kk'}{(k+k')^2} \right)
\]
and $E(m)$ is a complete elliptic integral of the second kind \cite{abr}. The angular integrals $I_m(k,k')$ for $m>2$ can be calculated in a similar manner. 

The Scr\"odinger equations \eqref{dox10} for each individual $m$ define the spectrum (eigenfunctions and eigenenergies expressed through $k_0$) of fractional hydrogen atom for that particular $m$ and all $n\geq m$. That is to say, the equation for $m=0$ \eqref{dox10} 
\begin{eqnarray}
\left(k^\mu+k_0^\mu \right)\psi_{n0\mu}(k) \nonumber \\
+\frac{1}{\pi}\int_0^\infty I_0(k,k')\psi_{n0\mu}(k')k'dk'=0 \label{m0}
\end{eqnarray}
(where $ I_0(k,k')$ is defined by Eq. \eqref{dox12}) determines the eigenfunctions $\psi_{00\mu}$ (ground state),
$\psi_{10\mu}$, $\psi_{20\mu}$ etc.  In its turn, the equation for $m=1$
\begin{eqnarray}
\left(k^\mu+k_0^\mu \right)\psi_{n1\mu}(k)\nonumber \\
+\frac{1}{\pi}\int_0^\infty I_1(k,k')\psi_{n1\mu}(k')k'dk'=0, \label{m1}
\end{eqnarray}
($I_1(k,k')$ is defined by Ex. \eqref{ii1}) determines $\psi_{11\mu}$,  $\psi_{21\mu}$, $\psi_{31\mu}$ etc. 

The equation \eqref{dox10} and its particular cases \eqref{m0} and \eqref{m1} are important results of the present paper as they constitute a new representation of quantum hydrogen atom problem both in ordinary ($\mu=2$) and in fractional ($\mu<2$) cases. To the best of our knowledge, this representation has not previously been derived in the literature. As we shall see below, it is especially useful in fractional case as they give a tremendous alleviation of both numerical and variational problem solution. This is because they permit to avoid the account for angular variable in numerical solution (the expansion in infinite series over spherical harmonics is not good idea for numerical solution of spectral problem for integral equation) and make variational one much easier.

\section{Variational solution of the spectral problem}

Here we suggest the method of variational solution of the set of radial equations \eqref{dox10}, thereby obtaining the eigenenergies and eigenfunctions of the fractional hydrogen atom for all admissible $\mu$. For that we come back to the Schr\"odinger equation in terms of energy $E$, rewriting Eq. \eqref{dox10} in the form
\begin{eqnarray}
k^\mu\psi_{nm\mu}(k) +\frac{1}{\pi}\int_0^\infty I_m(k,k')\psi_{nm\mu}(k')k'dk'=\nonumber \\
E\psi_{nm\mu}(k). \label{ursh1}
\end{eqnarray}
As the integration over $\varphi'$ has been already accomplished in the second term of Eq. \eqref{ursh1}, the integration over $\varphi$ reduces merely to multiplication by $2\pi$.   Multiplying both parts of \eqref{ursh1} by $2\pi \psi_{nm\mu}(k)$ (no need for complex conjugation as radial functions are real) and integrating over $kdk$, we obtain following variational functional
\begin{eqnarray}
E_{var}(k_0)=2\pi \int_0^\infty k^\mu \psi_{nm\mu}^2(k)kdk \nonumber \\
+ 2\int_0^\infty I_m(k,k')\psi_{nm\mu}(k,k_0) kdk \nonumber \\
\times \int_0^\infty \psi_{nm\mu}(k',k_0)k'dk'.\label{varfun}
\end{eqnarray}
The variational problem is now formulated as follows. For each state, characterized by two quantum numbers $m$ and $n$, substitute properly orthonomalized function  $\psi_{nm\mu}$ into 
$E_{var}(k_0)$ \eqref{varfun} and minimize the obtained function over $k_0$. This procedure will give the approximate eigenstate $E_{min}$ for given $n$ and $m$ as a function of L\'evy index $\mu$. To do so, we need to construct the orthonormal set of trial wave functions. For that we use the corresponding basis for $\mu=2$ \cite{pp} and modify it to our case of arbitrary $\mu<1$. 

The basis for $\mu=2$ reads \cite{pp}
\begin{eqnarray}
\psi_{nm}({\bf k})=C_{nm}f(k)P_n^{|m|}(\cos \theta)e^{im\varphi},\label{xd1}\\
\cos \theta=\frac{k_0^2-k^2}{k_0^2+k^2}, \ f(k)=\left(\frac{2k_0}{k^2+k_0^2}\right)^{3/2},\nonumber
\end{eqnarray}
where $C_{nm}$ are normalization constants and $P_n^{|m|}(z)$ are associate Legendre polynomials \cite{abr}. Note that the state \eqref{xd1} with each $n=0,1,2,...$ is $2n+1$ fold degenerate \cite{pp,kit}. In the case of functions \eqref{xd1}, the orthonormalization relation reads \cite{pp}
\begin{eqnarray}
\frac{1}{(2\pi)^2}\int f_w(k)\psi_{n'm'}^*({\bf k})\psi_{nm}({\bf k})d^2k=\delta_{mm'}\delta_{nn'}, \label{xd2}\\
f_w(k)=\frac{k^2+k_0^2}{2k_0^2}.\label{xd2a}
\end{eqnarray}
The choice of the weigh function $f_w(k)$ in the form \eqref{xd2a} is dictated by the demand that in ${\bf r}$ - space the wave function \eqref{xd1} should be normalized to unity. Below we shall see that for $\mu<1$ the orthonormalization condition \eqref{xd2} with weigh function \eqref{xd2a} will lead to divergent integrals. For that reason below we shall use different orthonormalization procedure. 

To construct our variational basis, we should first analyze the character of wave functions localization in momentum space. In other words, we should find the large $k$ asymptotics of the  $\psi_{nm\mu}({\bf k})$. This can be accomplished most conveniently based on the radial equation \eqref{dox10}. The large $k$ asymptotics of all $I_m(k,k')$ is $(k+k')^{-1}$. In this case, at $k \to \infty$ we can neglect $k_0^\mu$ in the first term of \eqref{dox10} and $k'$ in the denominators of integrands in the second term. This generates very simple algebraic equation 
$k^\mu \psi-A/(\pi k)=0$ for $\psi \equiv \psi_{nm\mu}(k \to \infty)$. Here $A=const=\int_0^\infty \psi(k')k'dk'$. The solution of the above algebraic equation gives the desired asymptotics
\begin{equation}\label{xd3}
 \psi_{nm\mu}(k \to \infty) \sim \frac{A}{k^{\mu+1}}.
\end{equation}
Based on Eq. \eqref{xd3}, it is reasonable to choose 
\begin{equation}\label{xd4}
 f_\mu(k)=\left(\frac{2k_0}{k^2+k_0^2}\right)^{\frac{\mu+1}{2}},
\end{equation}
which at $\mu=2$ gives the result \eqref{xd1}. It can be shown that the weigh function \eqref{xd2a} in our case modifies as 
\begin{equation}\label{xd5}
f_{w \mu}(k)=\left(\frac{k^2+k_0^2}{2k_0^2}\right)^{\mu-1},
\end{equation}
(once more, at $\mu=2$ we recover the result \eqref{xd2a}) so that the integrals like \eqref{xd2} become divergent at $\mu<2$. That is why in subsequent calculations we will not use the normalization \eqref{xd2} with weigh function. Rather, we shall use the "symmetric" normalization \eqref{qmm1} with corresponding modification of the basic functions  \eqref{xd1}. Namely, we substitute the coefficients of Legendre polynomials in Eq.\eqref{xd1} by the unknown ones, which will be found from the orthogonality conditions.

The ground state ($n=0$) is not degenerate and corresponding Legendre polynomial $P_0^0=1$. This implies that ground state wave function is simply proportional to $f_\mu(k)$ \eqref{xd4}. With respect to normalization condition \eqref{qmm1}, we have 
\begin{equation}\label{gswf}
\psi_{00\mu}(k)=\sqrt{\frac{\mu}{\pi}}\frac{k_0^\mu }{(k^2+k_0^2)^{\frac{\mu+1}{2}}},
\end{equation}
where $k_0$ is now a variational parameter.

The first excited state ($n=1$) is three-fold degenerate at $\mu=2$ so that there are three distinct functions $\psi_{10\mu}(k)$, $\psi_{11\mu}(k)e^{i\varphi}$, $\psi_{1,-1\mu}(k)e^{-i\varphi}$. It is well known that at $\mu=2$ the functions $\psi_{11\mu}(k)$ and $\psi_{1,-1\mu}(k)$ are similar to each other. Our analysis shows that at $\mu<1$ any appropriate choice (from the point of view of orhogonality, see below) of these functions gives the same energy. This means that it is reasonable to choose them equal to each other also for $\mu<1$. At the same time, below we shall see that the energies, corresponding to $\psi_{10\mu}(k)$ and $\psi_{11\mu}(k)$ are different, which implies that fractional Laplacian lifts the angular quantum number degeneracy of the hydrogen atom. We choose the functions  $\psi_{10\mu}(k)$ and $\psi_{11\mu}(k)$ in the form
\begin{eqnarray}
\psi_{10\mu}(k)=\sqrt{\frac{\mu+2}{\pi}}\ \frac{k_0^\mu(k_0^2-\mu k^2)}{(k_0^2+k^2)^\frac{\mu+3}{2}},\label{fi10}\\
\psi_{11\mu}(k)=-\sqrt{\frac{(\mu+1)(\mu+2)}{\pi}}\frac{k k_0^{\mu+1}}{(k_0^2+k^2)^\frac{\mu+3}{2}}.\label{fi11}
\end{eqnarray}
The coefficient $\mu$ in front of $k^2$ in the numerator of expression \eqref{fi10} is chosen from the orthogonality condition $\langle\psi_{10\mu}(k)\psi_{00\mu}(k)\rangle=0$, where $\langle...\rangle=2\pi \int_0^\infty...kdk$. Note that functions with different $m$'s are already orthogonal because of relation
\begin{equation}\label{ortug}
\int_0^{2\pi}e^{i(m-m')\varphi}d\varphi=2\pi \delta_{mm'}
\end{equation}
so that there is no need to orthogonalize the functions with different angular quantum numbers $m$. We only need to orthogonalize the functions with different $n$.

There are five distinct functions for the second excited state: $\psi_{20\mu}(k)$, $\psi_{2,\pm 1\mu}(k)e^{\pm i\varphi}$, $\psi_{2,\pm 2\mu}(k)e^{\pm 2i\varphi}$. Similar to the case $n=1$, the radial parts of the functions $\psi_{2,\pm 1\mu}(k)$ and $\psi_{2,\pm 2\mu}(k)$ are equal to each other, i.e. $\psi_{21\mu}(k)=\psi_{2,-1\mu}(k)$ and $\psi_{22\mu}(k)=\psi_{2,-2\mu}(k)$. We construct them in the form
\begin{eqnarray}
\psi_{20\mu}(k)=\sqrt{\frac{\mu+4}{\pi}}\ \frac{k_0^\mu(k_0^4+Bk^2k_0^2+Ck^4)}{(k_0^2+k^2)^\frac{\mu+5}{2}},\label{fi20}\\
\psi_{21\mu}(k)=-\sqrt{\frac{2(\mu+2)(\mu+4)}{\pi}}\frac{kk_0^{\mu+1}(k_0^2-Dk^2) }{(k_0^2+k^2)^\frac{\mu+5}{2}},\label{fi21}\\
\psi_{22\mu}(k)=\sqrt{\frac{(\mu+2)(\mu+3)(\mu+4)}{2\pi}}\frac{k^2 k_0^{\mu+2}}{(k_0^2+k^2)^\frac{\mu+5}{2}}.\label{fi22}
\end{eqnarray}
The coefficients $B$, $C$ and $D$ are found from orthogonality conditions $\langle\psi_{20\mu}(k)\psi_{00\mu}(k)\rangle=0$, $\langle\psi_{20\mu}(k)\psi_{10\mu}(k)\rangle=0$ and 
$\langle\psi_{21\mu}(k)\psi_{11\mu}(k)\rangle=0$, which yields
\begin{equation}\label{kofi2}
B=-2(\mu+1),\ C=\frac 12 \mu(\mu+1),\ D=\frac 12 (\mu+1).
\end{equation}
Presented construction shows our algorithm of orthogonalization. Namely, based on functions for $\mu=2$ \eqref{xd1}, we choose the functions for $\mu<2$ and given $n$ as the polynomials (in numerators) with unknown coefficients. These coefficients are determined from the corresponding orthogonality conditions similar to the above cases of $n=1$ and 2. Such procedure can be continued for arbitrary $n$, which gives a prescription to construct the arbitrary excited state.

Substitution of constructed wave functions to the expression for $E_{var}(k_0)$ \eqref{varfun} generates the expression 
\begin{equation}\label{gds5}
E_{var}(k_0)=k_0^\mu  {\mathcal J}_{1 nm\mu}-k_0 {\mathcal J}_{2nm\mu},
\end{equation}
which has the same structure for all quantum numbers $m$ and $n$. That is to say that the integrals ${\mathcal J}_{1 nm\mu}$	(coming from the first term in right-hand side of \eqref{varfun}) and ${\mathcal J}_{2 nm\mu}$ (coming from the second term in right-hand side of \eqref{varfun}) are independent of $k_0$. This, in turn, implies very simple procedure of finding of the energy \eqref{gds5} extremum, which yields
\begin{eqnarray}
k_{0\ extr}(\mu)=\left(\frac{{\mathcal J}_{2nm\mu}}{\mu {\mathcal J}_{1nm\mu}}\right)^{\frac{1}{\mu-1}},\label{gds6a} \\
E_{extr}(\mu)=k_{0\ extr}(\mu){\mathcal J}_{2nm\mu}\left(\frac{1}{\mu}-1\right), \label{gds6b}
\end{eqnarray}
where we suppress indices $m$ and $n$ in $k_{0\min}$ and $E_{min}$ for clarity. It is seen that point $\mu=1$ is peculiar for this procedure. Really, at $\mu=1$ we have that $E_{var}(k_0) \sim k_0$, i.e. is a linear function having neither maximum nor minimum. 

To understand the character of the above extremum for different $\mu$ we use the ordinary prescription of finding the second derivative $E_{var}''$ of the energy \eqref{gds5} in the extremum point \eqref{gds6a}. For $E_{var}''>0$ we have minimum, while for $E_{var}''<0$ we have maximum. The expression for second derivative in the point \eqref{gds6a} reads
\begin{equation}\label{gds7}
E_{var}''(k_{0\ extr})=\mu(\mu-1)k_{0\ extr}^{\mu-2} {\mathcal J}_{1nm\mu}
\end{equation}
It can be shown that the integrals ${\mathcal J}_{1,2\ nm\mu}$ are positive for all admissible $n$, $m$ and $\mu$. Hence $k_{0\ extr}(\mu)>0$ and $E_{var}''(k_{0\ extr})>0$ at $\mu>1$ and $E_{var}''(k_{0\ extr})<0$ at $\mu<1$, which shows that minimal solution of our variational problem exists at $\mu>1$ only. This is one of the main theoretical results of the paper. 

Our numerical solution of the integral equations \eqref{dox10} will confirm this conclusion. In other words, contrary to the other problems (like 1D potential well \cite{well} and/or harmonic oscillator \cite{osc}) of fractional quantum mechanics, the problem of fractional 2D hydrogen atom has solutions for $\mu>1$ only. This may be related to the non-1D character of the problem (hence more quantum numbers related to the number of degrees of freedom) under consideration. More detailed studies of this question (along with consideration of fractional 3D hydrogen atom) will be published elsewhere.

\section{Comparison of variational and numerical solutions}

The representation \eqref{dox10} can be used for the numerical solution of the Schr\"odinger equation for the problem under consideration. Performing substitutions $k/k_0=x$, $k'/k_0=y$ in it, we arrive at the following form of the 
Schr\"odinger equation, which is convenient for numerical solution
\begin{equation}\label{numm1}
\kappa_0\psi_{nm\mu}(x)=-\frac{1}{\pi}\frac{1}{x^\mu+1}  \int_0^\infty I_m(x,y)\psi_{nm\mu}(y)ydy,
\end{equation} 
where $\kappa_0=k_0^{\mu-1}$ so that the eigenenergy
\begin{equation}\label{numm2}
E=-\kappa_0^{\frac{\mu}{\mu-1}}.
\end{equation}
The functions $I_m(x,y)$ are defined by the expression \eqref{dox11}. The equation \eqref{numm1} is a linear spectral problem for the Fredholm integral equation \cite{pm}, which can be easily discretized with subsequent solution of the spectral problem for the obtained matrix. The eigenenergy of our problem $E$ is related to the eigenvalues $\kappa_0$ of the above matrix by the expression \eqref{numm2}. To obtain the satisfactory accuracy of the numerical solution of the eigenproblem \eqref{numm1}, we should typically diagonalize a 10000$\times$10000 matrix, which makes the task to be quite computer intensive.  

Our next step is to obtain the variational energies for the states $\psi_{nm\mu}$ as functions of L\'evy index $\mu$ by means of \eqref{gds6b} and compare them to numerical ones, obtained from \eqref{numm1}, \eqref{numm2}. Note that the forms of variational wave functions are dictated by the expressions \eqref{gswf} (ground state), \eqref{fi10}, \eqref{fi11} (first excited state) and \eqref{fi20} - \eqref{fi22} (second excited state) with respect to parameter $k_{0\ min}(\mu)$ \eqref{gds6a} as for $\mu>1$ the energy extremum corresponds to minimum. 

We have for the ground state ($n,m=0$) 
\begin{eqnarray}
{\mathcal J}_{1\ 00\mu}=\frac{\mu \Gamma^2\left(\frac{\mu}{2}\right)}{2\Gamma(\mu)}, \nonumber \\
{\mathcal J}_{2\ 00\mu}=\frac{8\mu}{\pi}\int_0^\infty \frac{xdx}{(x+x_1)(1+x^2)^\frac{\mu+1}{2}} \nonumber \\
\times \int_0^\infty \frac{x_1dx_1}{(1+x_1^2)^\frac{\mu+1}{2}}K\left(\frac{4xx_1}{(x+x_1)^2}\right), \label{vns1}
\end{eqnarray}
where $\Gamma(z)$ is gamma-function \cite{abr}. 

\begin{figure}
\begin{center}
\includegraphics[width=1.1\columnwidth]{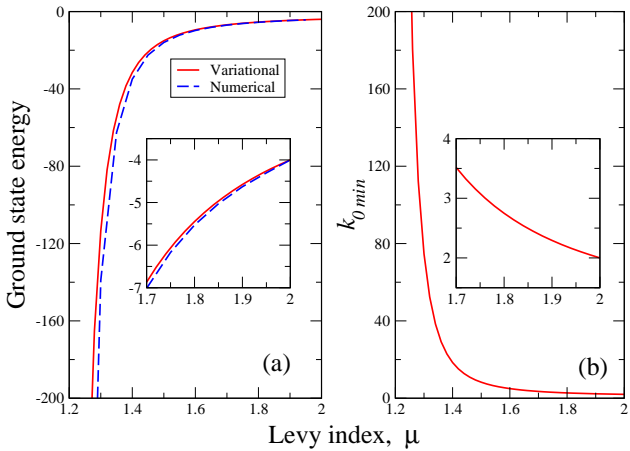}
\end{center}
\caption{Panel (a). Comparison of numerical (dashed line) and variational [Eq \eqref{gds6b} with respect to \eqref{vns1}, solid lines] ground state energies as functions of the L\'evy index $\mu$. Panel (b). Variational parameter $k_{0\ min}$ [Eq \eqref{gds6a} with respect to \eqref{vns1}] as a function of $\mu$. Insets to both panels show the ground state energy (panel (a)) and $k_{0\ min}$  (panel (b)) for $1.7<\mu<2$ to detail their behavior near
$\mu=2$ limit.}
\label{figg1}
\end{figure}

Now we consider the known case of "ordinary" (i.e. with normal Laplacian in the Schr\"odinger equation) 2D hydrogen atom, corresponding to $\mu=2$. In this case the solution of the problem is well studied (see \cite{kit} for coordinate space and \cite{pp} for momentum space) and its energy spectrum in our Rydberg units reads
\begin{equation}\label{spe2d}
E=-\frac{1}{(n+1/2)^2},\ n=0,1,2,3,...
\end{equation} 
At the same time, we have from \eqref{vns1} at $\mu=2$
\begin{equation}\label{vns2}
{\mathcal J}_{1\ 00\mu=2}=1,\ {\mathcal J}_{2\ 00\mu=2}=\frac{16}{\pi}\frac{\pi}{4}=4.
\end{equation} 
The value ${\mathcal J}_{2\ 00\mu=2}$ has been obtained numerically. Its explicit form is listed in Appendix \ref{chis}.
Substitution of \eqref{vns2} to the expressions \eqref{gds6a} and \eqref{gds6b} gives
\begin{equation}\label{vns3}
k_{0 min}(\mu=2)=2,\ E_{min}(\mu=2)=-4, 
\end{equation}
which reproduces the exact result \eqref{spe2d} for $n=0$ (ground state). 

\begin{figure}
	\begin{center}
		\includegraphics[width=1.1\columnwidth]{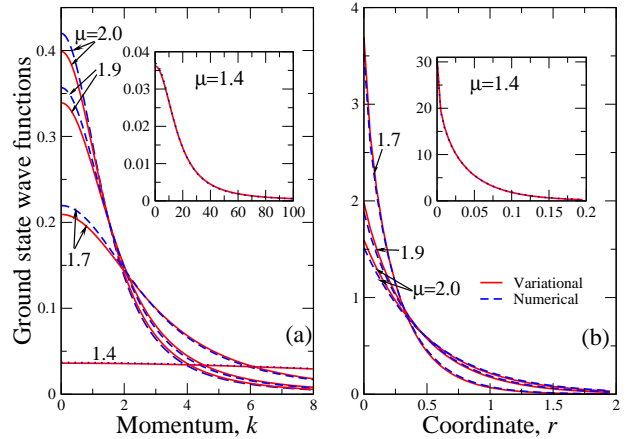}
	\end{center}
	\caption{Comparison of numerical (dashed line) and variational (Eq. \eqref{gswf}, solid line) ground state wave functions for different L\'evy indices $\mu$ (figures near curves). Panel(a) - momentum space, panel (b) - coordinate one. Insets to both panels detail the behavior of the function for $\mu=1.4$.}
	\label{figg2}
\end{figure}

The comparison of numerical and variational ground state energies is reported in Fig. \ref{figg1} (a). As usually for variational method, the variational curve (solid line) lies above the numerical one. This is because variational energy should be larger than its exact (in our case numerical) value \cite{land3}. It is seen that both numerical and variational ground state energies go to minus infinity as L\'evy index $\mu$ approaches 1. The inset to Fig. \ref{figg1} (a) shows the behavior of the energy at $\mu$ sufficiently close to 2, the latter case corresponds to ordinary quantum 2D hydrogenic problem. It is seen that at $\mu=2$ the numerical and variational solutions match which means that for ordinary quantum 2D hydrogen problem the variational method is exact, see also expression \eqref{vns3}. At $\mu<2$ the ground state energy diminishes, i.e. fractional hydrogen atom has lower energy then ordinary one. Panel (b) of Fig. \ref{figg1} the minimal radius of state (in momentum space) $k_{0\ min}$, found from our variational procedure. Inset shows once more the behavior at $\mu$ near 2. It is seen that the radius of state grows as $\mu$ diminishes from 2 to 1. At $\mu \to 1$ $k_{0\ min} \to \infty$.   

%%%%%%%%%%%%%%%%%%%%%%%%%%%%%%%%%%%%%%%%%%
\begin{table*}[t]
	\begin{tabular}{ccccccc}
		\hline \hline
		State label, $nm$ & 00 & 10 & 20 & 11 & 21 & 22 \\ \hline 
		$\mu=1.1$  &-8069.7 & -24.56 & -0.0465& -2.7317& -0.0356& -0.0074 \\ 
		$\mu=1.3$  &-139.8 & -7.5767& -0.1332& -0.5590& -0.1138& -0.0503\\ 
		$\mu=1.5$  &-15.983&-1.4516 &-0.1514& -0.4467& -0.1296&-0.0863 \\ 
		$\mu=1.7$  &-7.0138 & -0.6810& -0.1540&-0.4343&-0.1407& -0.1174 \\
		$\mu=1.8$  &-5.3659,& -0.5636 &-0.1541&-0.4372&-0.1463 & -0.1314\\ 
		$\mu=2.0$  &-3.9929&-0.4415& -0.1566& -0.4489&-0.1571&-0.1566\\ \hline \hline
	\end{tabular}
	\caption{Six lowest eigenvalues for angular momenta $m=0,1,2$ of the fractional quantum 2D hydrogen atom for different $\mu$, obtained numerically from matrix 10000$\times$10000 diagonalization. Maximal relative error is estimated to be 2\%. The energies for $m$ and $-m$ are the same, i.e. $E_{11}=E_{1,-1}$, $E_{21}=E_{2,-1}$, $E_{22}=E_{2,-2}$.}\label{yt}
\end{table*} 
%%%%%%%%%%%%%%%%%%%%%%%%%%%%%%%%%%%%%%%%%%%%%  

Substitution of the values $k_{0\ min}$ from Fig.  \ref{figg1} (b) into the expression \eqref{gswf} generates the ground state wave function for each $1<\mu \leq 2$. These functions are shown in Fig. \ref{figg2} along with their numerical counterparts. The functions in coordinate space are obtained by numerical inverse Fourier transformation \eqref{ftran2}. Very good coincidence between numerical and variational approaches is clearly seen. It can be shown that the average relative error between numerical and variational curves is almost the same for all $1<\mu \leq 2$ and equals approximately to 2\%. The curves for $\mu=1.4$ look like similar in the scales (both in momentum and coordinate space) of the figure as they are "anomalous" in the sense that they has very small amplitude and long spatial extension (inset to panel (a)) in $k$ space, while their behavior is opposite (large amplitude and short spatial extension, inset to panel (b)) in $r$ space. It is seen from Fig.\ref{figg2}(a) that in $k$ space as $\mu \to 1$ the wave function amplitude diminishes and it becomes progressively more extended. For instance, alredy for $\mu=1.4$ (inset to Fig.\ref{figg2}(a)) the wave function amplitude is around 0.04 (while at $\mu=2$, corresponding to "ordinary" case it ten times more) and its extension is up to $k \sim 100$ (while at $\mu=2$ it is up to $k \sim 8$). This means that at $\mu=1$ (limiting case of our solution existence) the wave function has zero amplitude and infinite spatial extension. This corresponds to Dirac $\delta$ function in the coordinate space. The corresponding coordinate dependences are reported in Fig.\ref{figg2}(b) and show opposite trend - at $\mu=2$ we have well-known exponential function $\exp(-2r)$ \cite{kit}, while at $\mu=1$ it is above Dirac $\delta$ function.

\begin{figure}
	\begin{center}
		\includegraphics[width=1.1\columnwidth]{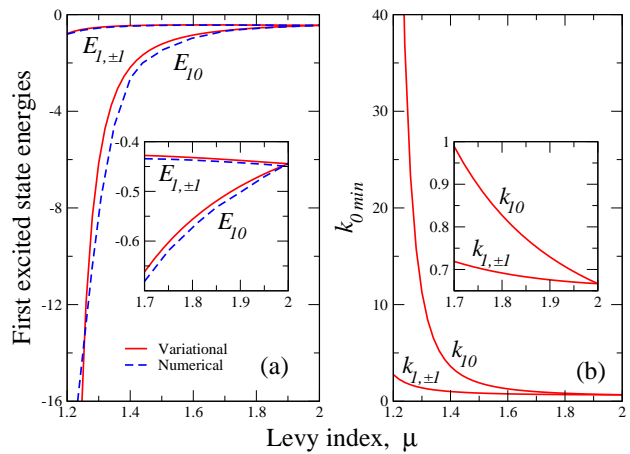}
	\end{center}
	\caption{Same as in Fig. \ref{figg1} but for the first excited state. Inset to panel (a) details the lifting of orbital degeneracy at $\mu<2$: while the energy $E_{10}$ diminishes monotonously, the energy $E_{1,\pm1}$ grows at $1.7<\mu<2$ and then decays much slower than $E_{10}$. Panel (b) shows the minimizing variational parameter
		as function of $\mu$. The difference between $k_{10}$ and $k_{1,\pm1}$ is clearly seen at $\mu<2$, which reflects the orbotal degeneracy lifting.}
	\label{figg3}
\end{figure}

Now we pass to the first excited state. Here a remarkable property of the "fractionalization" (the deviation of $\mu$ from 2 in our case) becomes visible. Namely, if at $\mu=2$ the 2D hydrogen atom energy is degenerate with respect to orbital quantum number $m$ (see Eq. \eqref{spe2d}), this degeneracy is lifted at $\mu<2$. Our analysis shows that this is the case for higher excited states also. The $\mu$ dependence of first excited state energies $E_{10}$ and $E_{1,\pm 1}$ (first lower index corresponds to principal quantum number $n$, while second to orbital number $m$) is reported in Fig.\ref{figg3}. Once more, good coincidence of numerical and variational energies is clearly seen in  Fig.\ref{figg3}(a). This is especially true for $E_{1,\pm 1}$, where numerical and variational curves are very close (less then 1\%) to each other at $1.2\leq \mu \leq 2$. For $E_{10}$, the coincidence (although generally not bad) is better at $\mu$ close to 2 (see the inset to Fig.\ref{figg3}(a)), where numerical energy lies lower than variational one, as it should be. However, at $\mu \sim 1.3$ the numerical curve intersects variational and at $\mu \leq 1.3$ and starts to have higher energy than variational one. This is related to the fact that for excited states the errors of our numerical scheme become larger so that we should use more (then $10^4$) discretization steps, which takes very long times for corresponding matrices diagonalization. Panel (b) of Fig. \ref{figg3} shows that minimizing momenta $k_0$ for the states $10$ and $1,\pm 1$ are different for $\mu<2$, yielding the resulting difference in the variational energies $E_{10}$ and $E_{1,\pm1}$. The same behavior of minimizing momenta is true for higher excited states.

Our numerical analysis of higher excited states shows that although we should choose larger number of discretization steps, the orbital degeneracy disappears at $\mu<2$. The main difference between higher excited states and first one is that there is (much for high $m$ and $n$) more states, which cease to be degenerate at $\mu<2$. Really, if for $n=1$, we have two states \eqref{fi10}, \eqref{fi11} (see also Fig. \ref{figg3}), for $n=2$ we have three states \eqref{fi20}, \eqref{fi21}, \eqref{fi22}, and for arbitrary $n$ we have $n+1$ states with respect to degeneracy (lifted in nonzero external magnetic field as it is related to time inversion symmetry \cite{land3}) $m \leftrightarrow -m$.
The selective results for the ground and first two excited states are summarized in the Table \ref{yt}. It is seen that the states 20, 21 and 22 have indeed the different energies at $\mu<2$. Also, if the ground state energy as well as those for the states 10 and 11 go to minus infinity as $\mu \to 1$, the energies for the states with $m=2$ do not. Rather, they go to zero or at least to very small values.  The point is that in the vicinity of $\mu=1$ our numerical method becomes unstable so that we should use more exquisite approaches in this region. The numerical calculations of the energies for higher $n$ show that they behave like those for $n=2$, i.e. grow from $\mu=2$ up to $\mu=1.05$, finishing at some small values. The energies for different $m$ are different for $\mu<2$, i.e. the orbital degeneracy is lifted.  This permits us to assert that the orbital degeneracy is lifted for 2D fractional quantum mechanical hydrogen atom.  

\section{Conclusions}

Two-dimensional fractional quantum-mechanical hydrogenic problem is a theoretical construction, which may have important physical realizations. For instance, it can describe phenomenologically (by means of fractional Laplacian introduction) the excitons in disordered semiconductor heterostructures and/or interfaces. In this paper, we have studied the spectral problem for 2D fractional quantum-mechanical hydrogen atom. Although 2D problem is much more complicated then 1D one, it turns out that in momentum space it admits approximate analytical solution by means of variational method. Moreover, the comparison of variational and numerical solutions to the problem permits us to establish two very important facts. First is related to fractional character of the problem. It tells us that the solution to the problem exists only at $1 < \mu \leq 2$. This is at odds with 1D fractional quantum mechanical problems of infinite potential well \cite{well} and oscillator \cite{osc}, where the solution exist for all admissible $0<\mu<2$. Second is the lifting of orbital degeneracy in fractional ($\mu<2$) case. This fact may have a profound influence on the described physical objects like excitons in semiconductors. This interesting question should be studied in more details. 

The immediate generalization of the considered problem is 3D fractional quantum hydrogen atom, which can be studied along the lines of present paper. Namely, we can readily pass to momentum space by means of symmetric 3D Fourier transform to obtain

\begin{equation}\label{mom3D}
\left(k^\mu+k_0^\mu \right)\Psi({\bf k})-\frac{1}{\pi^2}\int d^3k' \frac{\Psi({\bf k'})}{|{\bf k}-{\bf k'}|^2}=0,
\end{equation}
where ${\bf k}$ is now the 3D vector and parameter $k_0$ is again related to eigenenergy by Eq. \eqref{ek0}. The solution of 3D problem can be sought of in the form of a decomposition by radial $R_{nl\mu}(k)$ and angular parts
\begin{equation}\label{sol3D}
\psi_{nlm\mu}({\bf k})=R_{nl\mu}(k)Y_{lm}(\theta,\varphi),
\end{equation}
where one more quantum number $l$ (orbital quantum number \cite{land3}) appears. The spherical harmonics have usual form $Y_{lm}(\theta, \varphi)=P_l^{|m|}(\cos \theta)e^{im\varphi}$, where now $\theta$ is real (contrary to Eq. \eqref{xd1} for $\mu=2$, where $\theta$ was related to the modulus of $k$ in the spirit of stereographic projection \cite{pp}) apex angle. Our preliminary analysis shows that for $\mu \neq 2$ this problem also cannot be solved by the Fock stereographic projection \cite{fock} so that we should opt for variational and of course numerical methods. However, it is not yet clear if we can obtain the representation like \eqref{dox10} in 3D case. Latter fact complicates a lot the numerical and notably variational treatment of the problem. This makes the problem of 3D fractional quantum hydrogen atom to be much more arduous than 2D one.

Note that 3D fractional quantum-mechanical hydrogenic problem is also important for the description of Rydberg excitons (described by the quantum mechanical Kepler problem, see, e.g. \cite{anselm,ash}) in bulk semiconductors in the presence of disorder and other imperfections. Latter disorder, influencing the charge carrier (electron or hole in semiconductors) diffusion length, may lead to the excitonic spectrum, which cannot be described by the Schr\"odinger equation with ordinary Laplacian. Rather, fractional derivatives, which are usually responsible for "long tails", should be introduced in this case. For example, semiconducting perovskites like CsPbBr$_3$ are widely used in photovoltaics and have extremely long diffusion lengths \cite{stan}. The influence of disorder on their excitonic properties plays important role in spintronic (especially with Rashba spin-orbit coupling \cite{bibes}), optoelectronic and photovoltaic devises functionality. The explanation of this influence is still controversial, see, e.g. \cite{stan,yak}. One of the reasons is that the above perovskites have strong spin-orbit coupling, which alters the excitonic spectra \cite{stan,yak,tak}. This is closely related to the chaotic features in the spectra of excitons due to Rashba spin-orbit interaction \cite{she1,she2}. This suggests one more generalization of 2D and 3D  fractional quantum-mechanical hydrogenic problems. Namely, the spin-orbit interaction term can be added to corresponding fractional Schr\"odinger equation. In this case, the solution will be more sophisticated as the wave function will be spinor now \cite{she2,gri}, although the problem can become doable in the momentum space, where variational or direct diagonalization techniques \cite{she2} can be utilized. This problem turns out to be extremely important for above perovskite substances \cite{stan, yak,pho}, where chaos can disturb or even disrupt optoelectronic, spintronic and/or photovoltaic devices functionality. It had been shown \cite{she2} that the description by means of the "oridinary" (i.e. that with conventional Laplacian) Schr\"odinger equation do not show strong quantum chaotic features like non-Poissonian energy level statistics \cite{gutz,rei}. This may be related to the fact that proper description of such features is possible only within 2D and 3D excitonic models, containing fractional Laplacians. 

\begin{acknowledgements}
I am grateful to E. Ya. Sherman for fruitful discussions.
\end{acknowledgements}

\appendix
\section{Calculation of the integral $I_0(k,k')$ from the main text} \label{intt}
We calculate the integral \eqref{dox12}. The trigonometric identity  $\cos t=1-2\sin^2\frac t2$ and subsequent substitution $\theta=t/2$ reduces it to the form
\begin{eqnarray}
 I_0(k,k')=\frac{2}{\sqrt{A-B}}\int_{\varphi/2}^{\varphi/2-\pi}\frac{d\theta}{\sqrt{1+\frac{2B}{A-B}\sin^2 \theta}}=\nonumber \\
=  \frac{2}{\sqrt{A-B}}\biggl[F\left(\frac{\varphi}{2}-\pi, -\frac{2B}{A-B}\right)-\nonumber \\
-F\left(\frac{\varphi}{2}, -\frac{2B}{A-B}\right)\biggr],\ A=k^2+k'^2,\ B=2kk',\label{vbt1} 
\end{eqnarray}
where 
\begin{equation}\label{vbt2}
F(\varphi,m)=\int_0^\varphi \frac{d\theta}{\sqrt{1-m\sin^2\theta}}
\end{equation}
is incomplete elliptic integral of the first kind \cite{abr}. 

Making a substitution $\theta+\pi=z$ in the first integral, we obtain
\begin{widetext}
\begin{eqnarray}
&& I_0(k,k')=-\frac{2}{\sqrt{A-B}}\left[\int_0^{\varphi/2} \frac{dz}{\sqrt{1+\frac{2B}{A-B}\sin^2 z}}+\int_{\varphi/2}^\pi\frac{dz}{\sqrt{1+\frac{2B}{A-B}\sin^2 z}}\right]\equiv -\frac{2}{\sqrt{A-B}}\int_0^\pi \frac{dz}{\sqrt{1+\frac{2B}{A-B}\sin^2 z}}\equiv \nonumber \\
&&\equiv  -\frac{4}{\sqrt{A-B}}\int_0^{\pi/2} \frac{dz}{\sqrt{1+\frac{2B}{A-B}\sin^2 z}}\equiv -\frac{4}{\sqrt{A-B}}K\left(-\frac{2B}{A-B}\right),  \label{kpm0}
\end{eqnarray}
where $K(m)=F(\pi/2,m)$ is a complete elliptic integral of the first kind \cite{abr}. Although the negative argument is perfectly legitimate, it's a good idea to pass to the positive argument. This is accomplished as follows
\begin{eqnarray}
K(-m)=\int_0^{\pi/2}\frac{dx}{\sqrt{1+m\sin^2x}}=\int_0^{\pi/2}\frac{dx}{\sqrt{1+m-m\cos^2x}}=
\frac{1}{\sqrt{1+m}}\int_0^{\pi/2}\frac{d\theta}{\sqrt{1-\frac{m}{1+m}\cos^2x}}\nonumber \\
\equiv \frac{1}{\sqrt{1+m}}\int_0^{\pi/2}\frac{d\theta}{\sqrt{1-\frac{m}{1+m}\sin^2\theta}}\equiv \frac{1}{\sqrt{1+m}}K\left(\frac{m}{1+m}\right),\ 0<\frac{m}{1+m}<1. \label{kpm}
\end{eqnarray} 
\end{widetext}
To derive final expression \eqref{kpm}, we made a substitution $x=\pi/2-\theta$. Combining expressions \eqref{kpm0} and \eqref{kpm}, we obtain with respect to definitions of $A$ and $B$ \eqref{vbt1}
\begin{eqnarray}
 I_0(k,k')=-\frac{4}{\sqrt{A+B}}K\left(\frac{2B}{A+B}\right) \equiv\nonumber \\
 \equiv -\frac{4}{k+k'}K\left(\frac{4kk'}{(k+k')^2} \right),\label{vbt3}
\end{eqnarray}
which is the equation \eqref{dox12} from the main text.

Similar calculation for the complete elliptic integral of the second kind \cite{abr}
\begin{equation}\label{kpm1}
E(m)=\int_0^{\pi/2} \sqrt{1-m\sin^2\theta}d\theta
\end{equation} 
yields
\begin{equation}\label{kpm2}
E(-m)=\sqrt{1+m}\ E\left(\frac{m}{1+m}\right).
\end{equation}
The expression \eqref{kpm2} has been used to derive the expressions for $I_1(k,k')$ \eqref{ii1} and $I_2(k,k')$ \eqref{ii2} of the main text.

\section{Some exact results for the integrals ${\mathcal J}_{2\ nm\mu}$, obtained numerically} \label{chis}

Below we list the values of some of the integrals ${\mathcal J}_{2\ nm\mu=2}$ (although other $\mu$ values are also considered) in our variational approach, which we believe to be exact, although not tabulated in any known literature sources. The integrals below have been calculated numerically. But in their numerical evaluation we have purposely increased the precision and compared the result to known values, related to the number $\pi$, with the same precision. 

From ${\mathcal J}_{2\ 00\mu=2}$
\begin{eqnarray}
\int_0^\infty \frac{xdx}{(x+x_1)(1+x^2)^{3/2}}\nonumber \\
\times \int_0^\infty \frac{x_1dx_1}{(1+x_1^2)^{3/2}}K\left(\frac{4xx_1}{(x+x_1)^2}\right)=\frac{\pi}{4}.\label{az1}
\end{eqnarray}
From ${\mathcal J}_{2\ 00\mu=1}$
\begin{eqnarray}
\int_0^\infty \frac{xdx}{1+x^2}\nonumber \\
\times \int_0^\infty \frac{x_1dx_1}{(x+x_1)(1+x_1^2)}K\left(\frac{4xx_1}{(x+x_1)^2}\right)=\frac{\pi^3}{8}.\label{az2}
\end{eqnarray}
Many other numerically exact integrals can be extracted for higher $n$ and $m$.


\begin{thebibliography}{99}

\bibitem{bvb} M. Bibes, J. E. Villegas, and A. Barthel\'emy, Adv. Phys. {\bf 60}, 5 (2011).

\bibitem{as1} M. A{\ss}mann, J. Thewes, D. Fröhlich, and M. Bayer, Nat. Mater.
{\bf 15}, 741 (2016).

\bibitem{as2} J. Heck\"otter, M. Freitag, D. Fr\"ohlich, M. A{\ss}mann, M. Bayer, 
M. A. Semina, and M. M. Glazov, \prb {\bf 96} 125142 (2017).

\bibitem{ohm} A. Ohtomo and H. Y. Hwang, Nature (London) {\bf 427}, 423 (2004).

\bibitem{du1} V. A. Stephanovich and V. K. Dugaev, \prb {\bf 93}, 045302 (2016).

\bibitem{cells} M. Saliba, S. Orlandi, T. Matsui, S. Aghazada, M. Cavazzini,
J.-P. Correa-Baena, P. Gao, R. Scopelliti, E. Mosconi and
K.-H. Dahmen, et al., Nat. Energy {\bf 1}, 15017 (2016).

\bibitem{nanolas} H. Zhu, Y. Fu, F. Meng, X. Wu, Z. Gong, Q. Ding, M. V. Gustafsson, 
M. T. Trinh, S. Jin and X. Y. Zhu, Nat. Mater.  {\bf 14}, 636 (2015).

\bibitem{anselm} A.I. Anselm {\em{Introduction to semiconductor theory}} (Englewood Cilffs, N.J. : Prenctice-Hall, 1981).

\bibitem{ash} N.W. Ashkroft and N. D. Mermin, {\em{Solid State Physics}} (Harcourt,
New York, 1976).

\bibitem{she1} V.A. Stephanovich and E. Ya. Sherman  Phys. Chem. Chem. Phys.
{\bf 20}, 7836 (2018).

\bibitem{she2}  V.A. Stephanovich, E. Ya. Sherman, N.T. Zinner and O.V. Marchukov 
\prb {\bf 97}, 205407 (2018).

\bibitem{br} Yu. A. Bychkov and E. I. Rashba, JETP Lett., {\bf 39}, 78 (1984).

\bibitem{abr}  {\em{Handbook of Mathematical Functions}}, edited by M.
Abramowitz and I. Stegun (Dover, New York, 1972).

\bibitem{laskin} Nick Laskin  {\em{Fractional Quantum Mechanics}} (World Scientific, Singapore, 2018).

\bibitem{pp} D.G.W. Parfitt and M.E. Portnoi Journ. Math. Phys. {\bf 43}, 4681 (2002).

\bibitem{land3} L. D. Landau and E. M. Lifshits, {\em{Quantum Mechanics. Nonrelativistic Theory}} (Pergamon Press, Oxford, 1995).

\bibitem{kit} X.L. Yang, S.H. Guo, F.T. Chan, K.W. Wong, W.Y. Ching \pra {\bf 43}, 1186, 1991.

\bibitem{fock} V.A. Fock  Z. Phys. {\bf 98}, 145 (1935).

\bibitem{levy1} M. L\'evy Proc. Royal Soc. of London. Series A {\bf 204}, 145 (1950).

\bibitem{well} E.V. Kirichenko, P. Garbaczewski, V. A. Stephanovich and M. \.Zaba 
 \pre {\bf 93}, 052110 (2016).

\bibitem{osc}  E.V. Kirichenko and  V. A. Stephanovich  \pre {\bf 98}, 052127 (2018).

\bibitem{pm} A. D. Polyanin and A. V. Manzhirov, {\em{Handbook of Integral
Equations}} (Chapman \& Hall/CRC Press, Boca Raton-London, 2008).

\bibitem{stan} S.D. Stranks \& H.J. Snaith Nature Nanotechnology {\bf 10}, 391 (2015).
S.D. Stranks {\em{et al}} Science {\bf 342}, 341 (2013).

\bibitem{bibes} M. Bibes, J. E. Villegas and A. Barth\'el\'emy, Adv. Phys., {\bf 60}, 5 (2011).

\bibitem{yak} D. Canneson, E. V. Shornikova, D. R. Yakovlev, T. Rogge,
A. A. Mitioglu, M. V. Ballottin, P. C. M. Christianen, E. Lhuillier,
M. Bayer and L. Biadala Nano Lett., {\bf 17}, 6177 (2017).

\bibitem{tak} M. Fu, Ph. Tamarat, H. Huang, J. Even, A. L. Rogach, \& B. Lounis
Nano Lett. {\bf 17} 2895 (2017).

\bibitem{gri} C. Grimaldi, \prb {\bf 77}, 113308 (2008).

\bibitem{pho} W. S. Yang, J. H. Noh, N. J. Jeon, Y. C. Kim, S. Ryu, J. Seo and
S. I. Seok, Science, {\bf 348}, 1234 (2015).

\bibitem{gutz} M. C. Gutzwiller, {\em{Chaos in Classical and Quantum Mechanics}} 
(Springer-Verlag, New York, 1990).

\bibitem{rei} L. E. Reichl, {\em{The Transition to Chaos. Conservative Classical
Systems and Quantum Manifestations}} (Springer-Verlag, New York, 2nd edn, 2004).
\end{thebibliography}
\end{document}